\documentclass[preprint,aps,prb,showpacs,preprintnumbers]{revtex4}

\usepackage{amssymb}
\usepackage{amsmath}
\usepackage{dcolumn}
\usepackage{bm}
\usepackage{graphicx}
\usepackage{verbatim}
\usepackage{color}
\usepackage{textcomp}

\begin{document}

\preprint{AIP/123-QED}

\title[Sample title]{Electrolyte gate dependent high-frequency measurement of graphene field-effect transistor for sensing applications}

\author{W. Fu}
\email{Wangyang.Fu@unibas.ch; Christian.Schoenenberger@unibas.ch}
\author{M. El Abbassi,$^{\dag}$~T. Hasler}
\author{M. Jung}
\author{M. Steinacher}
\author{M. Calame}
\author{C. Sch{\"o}nenberger$^{\ast}$}
\affiliation{Department of Physics, University of Basel, Basel,
Switzerland} \affiliation{$^{\dag}$Ecole Normale
Sup$\acute{e}$rieure, International Center of Fundamental Physics,
Paris, France}

\author{G. Puebla-Hellmann}
\author{S. Hellm\"{u}ller}
\author{T. Ihn}
\author{A. Wallraff}
\affiliation{Department of Physics, ETH Z{\"u}rich, Z{\"u}rich,
Switzerland}

\date{\today}
\definecolor{darkgreen}{rgb}{.004,.68,.31}
\begin{abstract}

We performed radio\-frequency (RF) reflectometry measurements at
$2-4$\,GHz on electrolyte-gated graphene field-effect transistors
(GFETs) utilizing a tunable stub-matching circuit for impedance
matching. We demonstrate that the gate voltage dependent RF
resistivity of graphene can be deduced even in the presence of the
electrolyte which is in direct contact with the graphene layer.
The RF resistivity is found to be consistent with its DC
counterpart in the full gate voltage range. Furthermore, in order
to access the potential of high-frequency sensing for
applications, we demonstrate time-dependent gating in solution
with nanosecond time resolution.

\end{abstract}

\maketitle


Owing to its atomically thin structure and exceptional high
mobilities,~\cite{Novoselov04, Novoselov05, DasSarma2010} graphene
is potentially well suited to radio\-frequency (RF) applications.
This prospect is reinforced by the relative openness of the
RF-electronics industry to new materials without the requirement
of a high on/off current ratio, which limits the application of
graphene for digital applications.~\cite{Schwierz2010} Much of the
research conducted so far on graphene RF transistors has focused
on the cut-off frequency, $f_{T}$, which is the highest frequency
at which a field-effect transistor (FET) is useful in RF
applications.~\cite{Lin2010, Lin2010-2, Pallecchi2011,
Liao2010,Dean2010, Wang2011} For instance, graphene FETs (GFETs)
with an intrinsic cut-off frequency of $f_{T}=100-300$\,GHz have
been demonstrated,~\cite{Lin2010-2, Liao2010} which are superior
to the best silicon MOSFETs with similar gate lengths. Graphene
full-wave rectification and consequently frequency conversion with
high efficiency have also been demonstrated by making use of the
ambipolar conduction properties.~\cite{Wang2009, Wang2010} The
cyclotron motion of the charge carriers of graphene in a magnetic
field suggests further applications in non-reciprocal
components.~\cite{Sounas2011, Fallahi2012} Moreover, the microwave
properties of graphene antennas and transmission line have also
been investigated.~\cite{Perruisseau-Carrier2012, Dragoman2011}

Recently, there is also growing interest in applying graphene RF
transistors to biochemical sensing
applications.~\cite{Iramnaaza2011, Le2012} But in spite of the rapid
advances in recent years, our understanding of the RF properties of
graphene, especially with regards to sensing in a liquid
environment, is still incomplete. Although it is known that
atomically thin large area graphene behaves as a wide\-band resistor
due to negligible skin effect and kinetic
inductance,~\cite{Skulason2011} it is difficult to measure the
device resistance directly at RF. This is due to the large shunt
capacitance in conventional back or top-gated graphene RF
transistors having a significant influence on the RF performance,
hindering (if not preventing) the extraction of the intrinsic
parameters of graphene.

In contrast to conventional oxide based back or top gating,
electrolyte gating can be used to tune the properties of GFETs
without shunting the propagating RF signal. This is because of the
unique frequency dependent properties of the electrolyte. At DC
and relatively low frequency $\alt10$\,MHz the ions in the
electrolyte can instantly respond to potential changes. Due to
ideal screening, an electrode in the electrolyte, even if placed
far away from the graphene surface, is very strongly coupled to
the graphene layer and therefore able to induce large changes in
the carrier concentration in graphene. The typical Debye screening
length in an ionic buffer solution at a concentration of $100$\,mM
is as small as $1$\,nm. In contrast, at high frequency $>
100$\,MHz the ions in the electrolyte start to lag behind the AC
electric field due to the viscosity of the
solution.~\cite{Bockris1966} As a result, the electrolyte behaves
as a pure dielectric at microwave frequency and the RF shunt gate
capacitance is negligible considering that the physical gate
electrode in the electrolyte can be placed far away from the
graphene samples.

In this work, we explore the RF properties of electrolyte-gated
GFETs by using an RF reflectometry technique. We demonstrated that
it is possible to deduce the gate dependent RF resistivity of
graphene at microwave frequencies by modeling the graphene sheet
as an $RC$ dissipative transmission line. The extracted RF
resistivity of graphene is found to be consistent with its DC
counterparts in the full gate voltage range. As a
proof-of-principle for high-speed sensing applications in liquids,
we demonstrate in addition nanosecond time-resolved measurements
in electrolytes. We believe that this work opens the avenue for
further research on a new generation of biochemical sensors. For
example, owing to its wide bandwidth and significantly reduced
1/$f$ noise at high frequencies, our graphene RF device will
enable ultra\-fast measurements with a good detection limit,
allowing to explore new phenomena and physics at the solid-liquid
interface.


In reflectometry the intensity of the reflected wave of
an RF signal is measured relative to the input signal yielding the
frequency dependent reflection coefficient $S_{11}$. This value
depends on the load impedance. The sensitivity to the load is
optimized if the load impedance matches the characteristic
impedance of the transmission line at $50$\,$\Omega$. In our
graphene devices, typical resistances are in the range of
$1-100$\,k$\Omega$, which is far off the matching value of the
transmission line. To achieve a better signal, we use the recently
developed approach~\cite{Hellmuller2012, Puebla-Hellmann2012}
based on stub-matching~\cite{Pozar2005}. Here, a stub is added to
the waveguide to realize an impedance matching circuit that
converts the large impedance of the graphene devices to a value
close to $50$\,$\Omega$. The stub-matching circuit is realized on
a printed circuit board (PCB, RO4003C) with coplanar waveguides
(CPWs) and ground planes made from a $40$\,$\mu$m Cu film with an
additional $3-6$\,$\mu$m Ni/$0.1$\,$\mu$m Au plating and
integrated with the graphene sample, see Fig.~1. Reflectometry
measurements are performed with a vector network analyzer (VNA)
connected via a bias tee to the PCB. The bias tee is needed for DC
measurements. All the components are placed at room temperature.
Impedance matching is achieved by shunting the CPW connecting to the
graphene sample by a stub placed at a distance of $d_{1}$ from the
sample (Fig.~1b). The stub itself is a CPW with a length $d_{2}$ and
is terminated by a varactor diode. The capacitance $C_{d}$ of the
diode can be tuned by applying a DC voltage $V_{d}$ through an
additional on-chip bias tee as depicted in Fig.~1a and 1b (dashed
red circle). To achieve impedance matching, both $d_1$ and $d_2$
have to be chosen close to $\lambda$/4, where $\lambda$ is the
wavelength of the $3$\,GHz carrier signal. The reflectometry setup
is first calibrated up to the input of the PCB using standard
procedures and commercial calibration kits. To calibrate the PCB
board, frequency dependent measurements were performed with and
without the graphene device.

Fig.~1c show the reflection coefficient $S_{11}$ versus frequency
$f$ at different diode voltages $V_d$ without a graphene device.
First, by tuning the capacitance $C_d$ of the diode, the best
matching is achieved at around $3$\,GHz with $V_d=$8.72\,V (green
line, Fig.~1c) where $S_{11}$ is lower than $-60$\,dB. At this
resonant point at matching we measure a full width at half maximum
of around $100$\,MHz, which places an upper limit on the setup
bandwidth. Second, by fitting the frequency-dependent $S_{11}$
curves using previously described calibration methods and
models,~\cite{Hellmuller2012,Puebla-Hellmann2012} we are able to
extract the electrical lengths of $d_{1}=13.92$\,mm and
$d_{2}=14.12$\,mm, values that are close to the design values
$d_{1}=13.45$\,mm and $d_{2}=13.75$\,mm. The extracted attenuation
constant of the CPW on the PCB is found to be $\alpha=5.15$\,dB/m
at a frequency of $3$\,GHz, which is comparable to $\alpha\sim
4.6$\,dB/m obtained from simulations of the PCB board (TXLINE
2003). After this calibration measurements, the graphene sample is
glued into an $8\times8$\,mm$^{2}$ recess at the end of the
$d_{1}$ CPW and connected with several bonding wires at both
sides. There are bonds from the $d_1$ stub to the graphene and on
the opposite side a series of bonds to the ground plane. The graphene
sample together with the bonding wires is then further
encapsulated to allow for measurements in a liquid environment.

Previously, we have demonstrated that high-mobility GFETs with
clean surfaces, and thus preserved chemical properties, can be
fabricated using large-scale chemical vapor deposition
(CVD).~\cite{Fu2011, Fu2013} These transistors can serve as an
ideal platform for biochemical sensing applications. In this
study, we explore the microwave properties of such GFETs. As shown
in Fig.~2a, we start with large area CVD graphene synthesized on
Cu.~\cite{Fu2011, Fu2013, Li2009} Then a uniform layer of epoxy is
put on a teflon substrate and a piece of graphene on copper is
carefully placed over it upside down. The graphene flake is in
direct contact with epoxy and on the opposite side the copper
layer faces air. The edges of the Cu surface are then manually
covered with PMMA as an easy way to obtain source and drain
electrodes after etching the Cu film in the middle. Due to the
good adhesion of the graphene flake on epoxy, the exposed Cu can
be etched away in an ammonium persulfate solution leaving the
intact graphene underneath. Afterwards, the PMMA can be carefully
peeled away to release the Cu electrodes. Such fabricated GFETs on
teflon substrates are placed at the end of the stub tuner and
connected with wire bonds. At the end of fabrication, an
additional epoxy step is applied to seal the Cu electrodes as well
as the bonding wires from liquid.


Fig.~2b shows the schematics and measurement circuits of such a
GFET device. The electrostatic potential in the electrolyte is
defined via a calomel reference electrode. This potential is
gating the graphene sample. Fig.~2c shows the dependence of the
measured DC electrical resistance $R$ of a GFET as a function of
the gate voltage applied to the calomel reference electrode
$V_{ref}$ (pink circle: forward, black star: backward sweep) in a
$100$\,mM KCl solution. Negligible hysteresis can be identified in
this transfer characteristics. This is an indication of the
reliability and reproducibility of the device and the
measurements.

After evaluating the DC transfer characteristics of the
electrolyte-gated GFET, we measured its frequency dependent
reflection coefficient $S_{11}$ when sweeping the liquid gate
voltage $V_{ref}$ from $-0.7$\,V to $0.1$\,V in steps of $0.1$\, V
in $100$\,mM KCl solution. For reasons of clarity, Fig.~3a shows
only a few curves measured at $-0.7$\,V (green dots), $-0.2$\,V
(black dots), and $0.1$\,V (blue dots). In order to fit the measured
curves, we introduce an $RC$ dissipative transmission line model.
This model treats the graphene sheet as a distributed $RC$ line
along which the RF signal propagates. We thereby neglect the
graphene inductance which yields a negligible contribution to the
impedance in our frequency range.~\cite{Skulason2011} Fig.~3a
(inset) depicts such a circuit. Here, $R'$ represents the resistance
of graphene and $C'$ the shunt capacitance both normalized per unit
length. An RF wave propagating along the $RC$ transmission line is
damped due the resistive component. At a distance $x$, the amplitude
of the RF signal will decrease by a factor of $exp[-\sqrt{\omega R'
C'/2}x]$, where $\omega$ is the angular frequency. The
characteristic propagation length over which the signal decays by a
factor of $1/e$ is then given by $L_{\lambda}=\sqrt{2/\omega R'
C'}$. The capacitance per unit length $C'$ is determined by the
dielectric materials leading to ground. As the ions in the liquid
cannot follow the RF field at $3$\,GHz the main part of the
capacitance is given by the substrate, and
since the substrate capacitance is much smaller than the quantum
capacitance $C_Q$,~\cite{Xia2009} the latter does not play a role at
RF frequencies.
In order to keep this capacitance low enough we have chosen to work
with teflon as a substrate material, hence,
$C'={\varepsilon_{o}}{\varepsilon_{teflon}}W/d$, with $d=1$\,mm the
thickness of the teflon substrate and $W=2.4$\,mm the width of the
graphene sheet. The water solution (electrolyte) on the upper
surface yields only a secondary contribution to $C'$ which is then
captured in the fitting by a slight increase in the effective
dielectric constant relative to the published value of
$\varepsilon_{teflon}$.

Having established a calibration up to the open end of the CPW, we
need to also measure the contact resistance $R_s$ and stray
capacitance $C_s$ of the bonding wires including the bonding pads.
These values are obtained by fitting similar curves as in Fig.~3a
after removal of the graphene in an O$_{2}$ plasma. Consequently,
there are only two fitting parameters for each curve in Fig.~3,
$C'$ and $R'$. $C'$ is related to an effective dielectric
constant, while we convert $R'$ into a sheet resistance or
resistivity $\rho$ given by $\rho=R'w$. The result of the fitting
is shown in Fig.~3a as dashed curves. It turns out that all curves
can be fitted with the same capacitance value $C'$. The extracted
effective $\varepsilon_{teflon}=2.4$ is found to be close to the
reported dielectric constant of teflon which is $2.1$. This
strongly supports the proposed model. We stress here that this
finding was reached by measurements carried out in both air and
liquid environments at different salt concentrations and pH values
(not shown).

The gate-dependent resistivity values $\rho_{RF}$ deduced from the
$S_{11}$ measurements are shown together with the DC measurements
$\rho_{DC}$ in Fig.~3b. As can be seen, there is a perfect match
between $\rho_{RF}$ and $\rho_{DC}$. This is in agreement with the
notion that the skin effect is negligible in ultra thin monolayer
graphene.~\cite{Dragoman2011, Skulason2011} The graphene RF
resistivity $\rho_{RF}$ ranges between $1$ and $5$\,k$\Omega$,
yielding a range of $0.4$ to $2$\,k$\Omega$/mm for $R'$. The
shunting capacitance per area
$C_{\square}=\varepsilon_{o}\varepsilon_{teflon}/d\sim2\times10^{-8}$\,F/m$^{2}$
yields $C'=0.5$\,aF/mm. This predicts a propagation length
$L_{\lambda}\sim 1-2.3$\,mm at $3$\,GHz, which is rather short.
When the graphene sheet is longer than $1$\,mm, which is the case
in the above device, the RF wave cannot propagate to the end, but
most of the signal is dissipated before. We therefore stress that it is
important to use a transmission line model and not only a fixed
two-terminal graphene resistance. The latter parameter can be
measured at low frequency. At high frequency the the
resistivity is the more natural physical quantity.

Having demonstrated the ability to measure the gate-dependence of
the graphene resistivity in an electrolyte, one may wonder how
large the attainable measurement bandwidth is. We therefore
performed time resolved measurements on the electrolyte-gated
GFETs. The used homodyne measurement setup is sketched in Fig.~4a.
A square wave at ~$2$\,MHz with a rise time of $7$\,ns is
generated by a pulse generator and applied to the electrolyte gate
of the GFET via a Au wire. The peak-to-peak voltage of the square
wave is $200$\,mV and its low level is adjusted to the $V_{CNP}$
of the GFET. At the same time, a $-30$\,dBm RF carrier signal
(here at $3.55$\,GHz) is applied to the input of the stub
tuner/GFET device via a directional coupler ($-20$\,dB coupling).
The resulting reflected signal exits the output port of the
directional coupler and is down-converted to DC using a mixer with
the carrier signal as local oscillator. Afterwards the signal is
low pass filtered, amplified and recorded by an oscilloscope at a
sampling rate of $1$\,GHz. Fig.~4b depicts the reference square
wave (applied to the electrolyte gate via a Au wire) and the
collected reflected signals for different KCl concentrations
ranging from $1$\,mM to $1$\,M. We measure a rise time of
$t_{rise}=50$\,ns in KCl solution for both $100$\,mM and $1$\,M
concentrations. In contrast, in case of the $10$\,mM KCl solution
the reflected signal starts to show a significant delay compared
to the $2$\,MHz square wave. At even lower concentration of
$1$\,mM concentration, the reflected signal is small and cannot
build up on this time scale. Additional measurements reveal rise
times of about $0.4$\,$\mu$s and $4$\,$\mu$s for a $10$\,mM and a
$1$\,mM concentrated solutions.

The rise time is plotted in the inset of Fig.~4b as a function of
ion concentration $c$. It can be quantitatively explained by looking
into the equivalent $RC$ circuits of the electrolyte-gated GFETs,
which consists of the interfacial capacitance $C_{I}$ and the series
resistances to which both the water solution and graphene device
contribute. First, the DC (or low frequency) interfacial capacitance
$C_{I}$ of an electrolyte-gated GFET can be modeled as two
capacitors in series.~\cite{Fu2013, Xia2009} One part is the quantum
capacitance of graphene, $C_{Q}$, which is gate voltage dependent
and has its minimum value at the charge neutral point (CNP). The
other part is the double layer capacitance, $C_{DL}$, which is
independent to the gate voltage. Here, in our case of moderate to
high ionic strength (from $1$\,mM to $1$\,M), the double layer
capacitance $C_{DL}$ is limited by the Stern layer capacitance
$C_{Stern}$ which is on the order of $0.2$\,F/m$^{2}$. This number
is over one order of magnitude larger than the quantum capacitance
of graphene $C_{Q}\sim$0.01 F/m$^{2}$ near the CNP at low carrier
density. Hence, $C_Q$ dominates the interface capacitance in the low
frequency regime.~\cite{Fu2013, Xia2009} Secondly, the resistance
between the water solution and the graphene flake is inversely
proportional to the KCl concentration, $R_{W}\propto1/c$. It can be
estimated to be about $10$\,k$\Omega$ to $10$\,M$\Omega$ for KCl
concentrations in the range of $1$\,mM to $1$\,M. As a comparison,
the resistance of our graphene flake is of the order of k$\Omega$.
This is up four to orders of magnitude lower than that of the water
solution $R_{W}$. Based on this discussion, the time constant of the
system can be estimated as $\tau=R_{W}C_{Q}\propto1/c$. This trend
satisfactorily explains the observed KCl concentration dependent
rise times of the electrolyte-gated GFET system seen in the inset of
Fig.~4 (grey line). A saturation of the time scale occurs at $\sim
50$\,ns for large ion concentrations. This is an extrinsic effect
mainly caused by the amplifier which (only) had a $10$\,MHz
bandwidth, limiting the response to $35$\,ns. The linear dependence
shows, however, that a $10$\,ns response time is feasible in a
highly concentrated buffer solution as used in biological studies.
This time scale would then also approach the bandwidth limtit of the
impedance matching circuit. The presented measurements demonstrate
the feasibility of achieving nanosecond time resolution in
measurements of the electrostatic potentials at the electrolyte
graphene surface.


In conclusion, this work represents a systematic study of the gate
voltage dependent characterization of large scale monolayer CVD GFET
at microwave frequencies in a liquid environment. Using the
reflect\-ometry technique, we measured the frequency dependent
reflection coefficient $S_{11}$ at different electrolyte gate
voltages. An $RC$ dissipative transmission line model is proposed to
extract the RF resistivity of GFETs after de-embedding, which is
found to follow its DC counterparts. The tunable transmission lines
realized by using high-mobility liquid-gated GFETs can serve as a
platform for a new generation of bio-chemical sensors. Potentially,
the wide bandwidth (100 MHz) offered by this high-frequency
measurements enables ultra fast measurements in liquid with $10$\,ns
time resolution.

\begin{acknowledgments}
The authors acknowledge funding from the Swiss Nano\-science
Institute (SNI),Swiss National Science Foundation, nanotera.ch,
NCCR-QSIT, ERC project QUEST, FP7 projects Symone and Graphene
Flagship. W.~Fu and M.~El~Abbassi contributed equally to this work.
\end{acknowledgments}


\begin{figure}
\includegraphics[width=140mm]{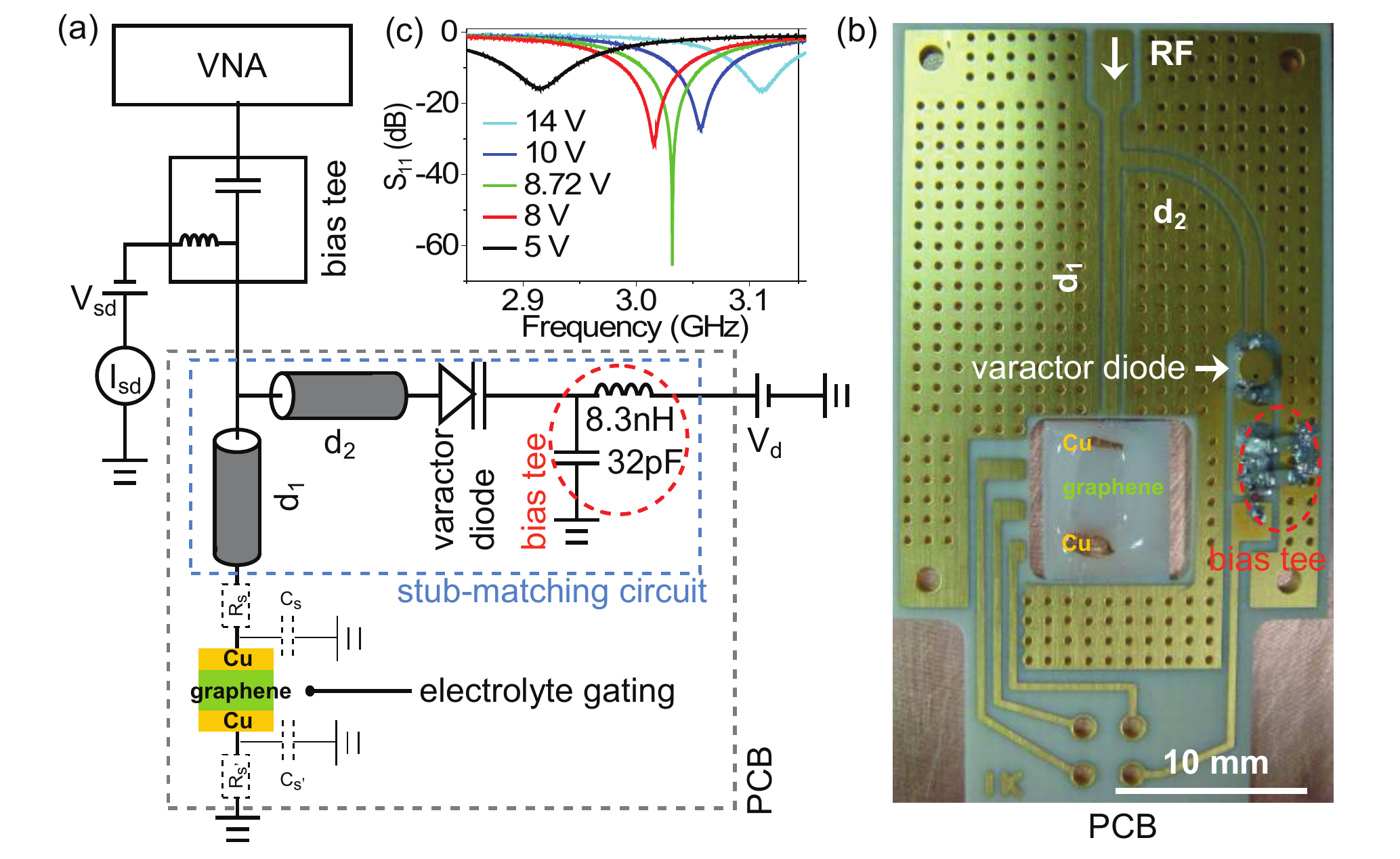}
\caption{\label{Figure1} (a) Experimental setup used to perform
the reflectometry measurements at carrier frequencies of 2-4 GHz.
(b) A picture of the PCB with a terminated waveguide ($d_{1}$) and
a shunted stub ($d_{2}$). The graphene sample is connected to the
$d_{1}$ waveguide via wire bonding and $d_{2}$ stub is terminated
by a varactor diode, after which two capacitors and one inductor
are placed as on-chip bias tee. (c) Open loop reflection
coefficient $S_{11}$ as a function of the carrier frequency $f$
measured under different varactor diode bias voltages $V_{d}$. }
\end{figure}

\begin{figure}
\includegraphics[width=140mm]{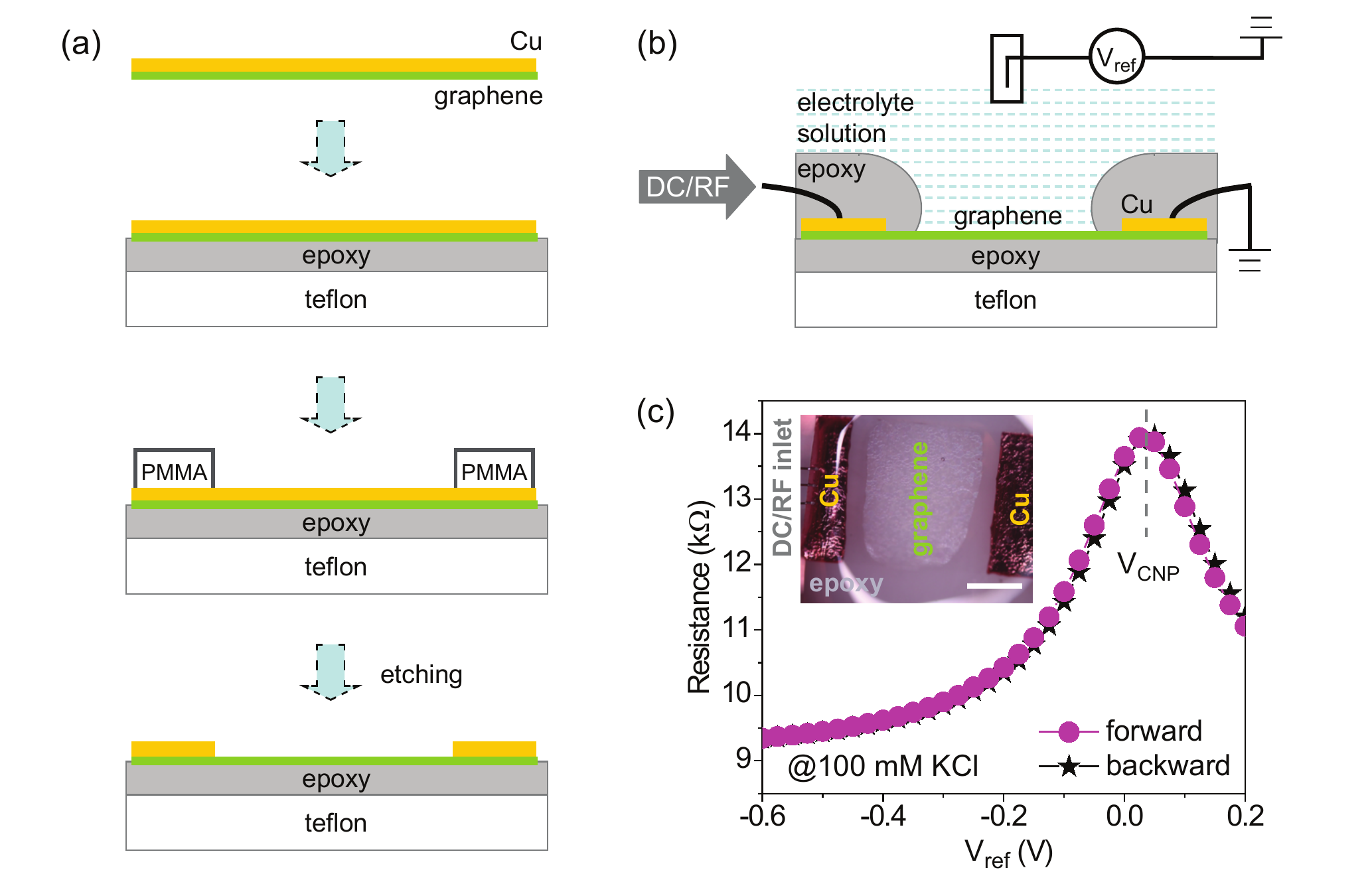}
  \caption{\label{Figure2}
  GFET fabrication and measurement scheme. (a) Process flow for transferring and patterning a
  CVD graphene sheet (green) with clean surface. (b) Schematic of a fabricated GFET with liquid
  sealing and the electrical circuitry of the electrolyte-gated GFET. The electrostatic potential
  in the solution was defined by a commercial calomel reference electrode as $V_{ref}$ or by a Au wire.
(c) Electrical resistance of a GFET as a function of the reference
potential $V_{ref}$ measured in $100$\,mM KCl solution with
negligible hysteresis. Inset, an optical image of the GFET under
test. Scale bar: 1 mm. }
\end{figure}

\begin{figure}
\includegraphics[width=160mm]{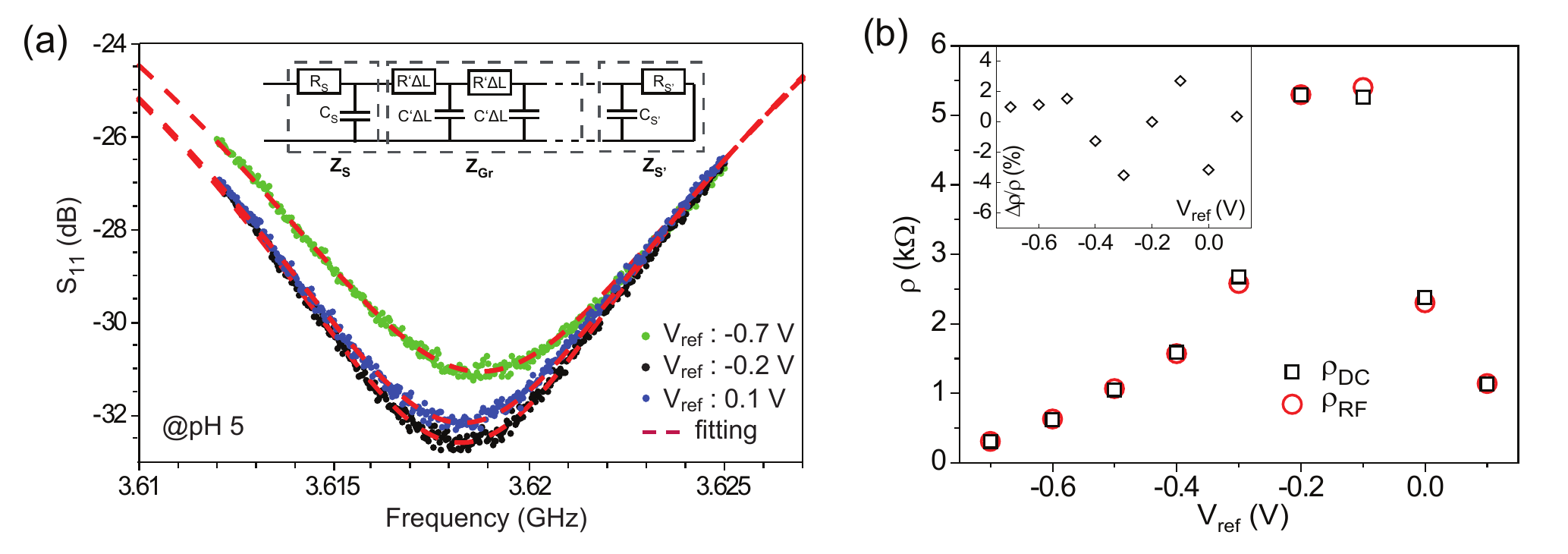}
  \caption{\label{Figure3} (a) The measured reflection spectra at different liquid gate voltages $V_{ref}$
  and the corresponding fitting curves. Inset: the $RC$ dissipative transmission line model. (b) The measured
  DC resistivity $\rho_{DC}$ (black squares) and the extracted small-signal RF resistance $\rho_{RF}$ (red circles) as a function of the applied
  liquid gate voltage $V_{ref}$. Inset: the relative deviation between $\rho_{DC}$ and $\rho_{RF}$ ($\Delta\rho/\rho:=(\rho_{RF}-\rho_{DC})/\rho_{DC}\times100~\%$) as a function of the
  applied liquid gate voltage $V_{ref}$.
}
\end{figure}

\clearpage

\begin{figure}
\includegraphics[width=160mm]{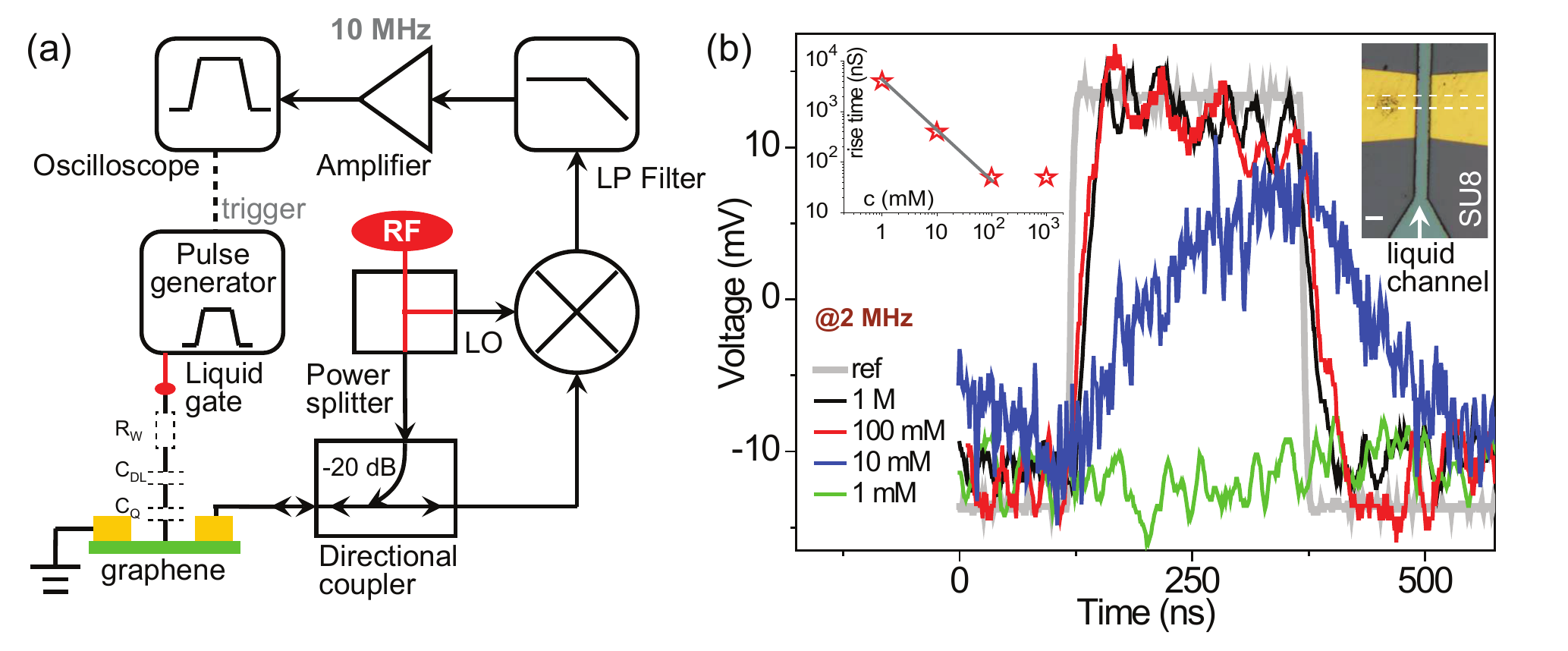}
  \caption{\label{Figure4} (a) Schematic of the time-resolved homodyne measurement setup. (b) Sensing responses
of the GFET to a $500$\,ns square wave reference signal with
$200$\,mV peak-to-peak voltage applied to the liquid gate
(normalized, grey line), in KCl solutions with concentration of
$1$\,mM(green line), $10$\,mM(blue line), $100$\,mM(red line) and
$1$\,M(black line). Inset, to the right: an optical image of the
GFET with SU-8 (a photoresist) liquid sealing. The dashed white
line is an indication of the graphene ribbon. Scale bar:
$10$\,$\mu$m. To the left: measured rise times as a function of
the KCl concentration $c$. The grey line indicates a linear 1/$c$
behavior.}
\end{figure}

\end{document}